\UseRawInputEncoding

\documentclass[showpacs,floats,floatfix,superscriptaddress,aps,pra]{revtex4-1}
\usepackage{amsfonts}
\usepackage{amssymb}
\usepackage{amsmath}
\usepackage{calc}
\usepackage{graphicx}
\usepackage{bm}

\usepackage[normalem]{ulem}
\usepackage{amsmath,amssymb}
\usepackage{multirow}
\usepackage{xcolor,soul}
\usepackage{srcltx}
\usepackage{hyperref,graphicx}

\def\be{ \begin{equation}}
\def\ee{ \end{equation}}
\def\bea{ \begin{eqnarray}}
\def\eea{ \end{eqnarray}}
\def\bse{ \begin{subequations}}
\def\ese{ \end{subequations}}
\def\bc{ \begin{center}}
\def\ec{ \end{center}}

\newcommand{\stef}[1]{{\color{black} #1}}

\begin{document}

\author{Stefano Longhi$^{*}$} 
\affiliation{Dipartimento di Fisica, Politecnico di Milano, Piazza L. da Vinci 32, I-20133 Milano, Italy}
\affiliation{IFISC (UIB-CSIC), Instituto de Fisica Interdisciplinar y Sistemas Complejos, E-07122 Palma de Mallorca, Spain}
\email{stefano.longhi@polimi.it}

\title{Phase transitions in a non-Hermitian Aubry-Andr\'e-Harper model }
  \normalsize

%\date{.}

%
\bigskip
\begin{abstract}
\noindent  
 The Aubry-Andr\'e-Harper model provides a paradigmatic example of aperiodic order in a one-dimensional lattice displaying a delocalization-localization
phase transition at a finite critical value $V_c$ of the quasiperiodic potential amplitude $V$.
 In terms of dynamical behavior of the system, the phase transition is
discontinuous when one measures the quantum diffusion exponent $\delta$ of wave packet spreading, with $\delta=1$ in the delocalized phase $V<V_c$ (ballistic transport), $\delta \simeq 1/2$ at the critical point $V=V_c$ (diffusive transport), and $\delta=0$ in the localized phase $V>V_c$ (dynamical localization).  However, the phase transition turns out to be smooth when one measures, as a dynamical variable, the speed $v(V)$ of excitation transport in the lattice, which is a continuous function of potential amplitude $V$ and vanishes as the localized phase is approached. Here we consider a non-Hermitian  extension of the Aubry-Andr\'e-Harper model, in which hopping along the lattice is asymmetric, and show that the dynamical localization-delocalization transition is discontinuous not only in the diffusion exponent $\delta$, but also in the speed $v$ of ballistic transport. This means that, even very close to the spectral phase transition point, rather counter-intuitively ballistic transport with a finite speed is allowed in the lattice. Also, we show that the ballistic velocity can increase as $V$ is increased above zero, i.e. surprisingly disorder in the lattice can result in an enhancement of transport.

\end{abstract}

\maketitle
      
\section{Introduction}

One-dimensional lattices with aperiodic order, i.e. displaying a long-range periodicity intermediate between ordinary periodic crystals and disordered
systems, provide fascinating models to study unusual transport phenomena in a wide variety of classical and quantum systems, 
ranging from condensed-matter systems to ultracold atoms, photonic and acoustic systems \cite{r0,r1,r2,r3,r4,r5,r6}.
Quasiperiodicity gives rise to a range of unusual behavior
including critical spectra, multifractal eigenstates, localization transitions at
a finite modulation of the on-site potential, and mobility edges \cite{R3,R4,R5,R6,r7,r8,r9,r10,r11,r12,r13,r14,r15,r16,kaz1,kaz2,kaz3,r17,r18,kaz4}. 
Typical dynamical variables that characterize single-particle transport are the largest propagation speed $v$ of excitation in the lattice, which is bounded (to form a light cone) for short-range hopping according to the Lieb-Robinson bound \cite{Lieb}, and the quantum diffusion exponent $ \delta$, which measures the asymptotic spreading of wave packet variance $\sigma^2(t)$ in the lattice according to the power law $\sigma^2(t) \sim t^{2 \delta}$. \stef{Such dynamical quantities are highly relevant in experiments, since they can be readily measured detecting the temporal spreading of an initially localized wave packet \cite{r1,r3,r5,r38}, and can thus provide insightful information about the underlying properties of the system.}
While there is not a one-to-one correspondence between spectral and dynamical properties of a quantum system \cite{uffa,uffa1}, a general rule of thumb is that absolutely continuous spectrum and extended states yield ballistic transport 
($v \neq 0$ and $\delta=1$), pure-point spectrum and exponentially-localized states usually yield  dynamical localization, i.e. suppression of transport and quantum diffusion ($v=0$ and $\delta=0$), while singular continuous spectrum and critical states yield diffusive (or anomalous diffusive) behavior ($v=0$, $0<\delta<1$). \stef{The localization and transport properties are clearly influenced by the strength and kind of disorder (either stochastic or incommensurate) of the system. While disorder is likely to contrast wave spreading, in certain models the opposite can occur, i.e. disorder can enhance spreading and transport \cite{cogl1,cogl2,cogl3,cogl4,Segev,cogl5,cogl6,cogl7,cogl8,cogl9}. In particular, evidence that weak disorder can  enhance propagation has been experimentally demonstrated in photonic quasicrystals \cite{Segev,cogl5}.}  \\ 
A paradigmatic physical example of a one-dimensional lattice with aperiodic order is provided by the Aubry-Andr\'e-Harper model \cite{r19,r20,r21},  also known as the almost Mathieu operator on a lattice in the mathematical literature \cite{r21a,r21b}. The Hamiltonian displays a Cantor-set energy spectrum  with a phase transition from extended states and absolutely continuous spectrum to exponentially localized states and pure point spectrum as the amplitude $V$ of the on-site quasi-periodic potential is increased above a critical value $V_c$ \cite{r21a,r21b}. In terms of dynamical behavior of a wave packet, measured by the exponent $\delta=\delta(V)$ of wave spreading, the phase transition is discontinuous since $\delta(V)=1$ for $V<V_c$ (ballistic transport), $\delta (V) \simeq 1/2$ at the critical point $V=V_c$ (almost diffusive transport), and $\delta(V)=0$ in the localized phase $V>V_c$ (dynamical localization) \cite{r13,r16}; see Fig.1(a).
However, in terms of the spreading velocity $v=v(V)$, defined as $v \sim \sigma(t)/t$, the phase transition turns out to be smooth, with $v(V)$ continuous function of potential amplitude $V$ and $v(V)=0$ for $V \geq V_c$ [see Fig.1(a)]. This result corresponds to physical intuition that in terms of dynamical evolution of the system the transition from ballistic wave packet spreading to dynamical localization, as the potential amplitude $V$ is increased above the critical value $V_c$, is a continuous process.\\
Recently, fresh and new perspectives on spectral localization, transport and topological phase transitions have been disclosed in non-Hermitian lattices, where complex on-site potentials or asymmetric hopping are phenomenologically introduced to describe system interaction with the surrounding environment \cite{r22,r23,r24,r25,r27,r28,r29,r30,r31,r32,r33,r34,r35,r36,r38,r39,r40,r41,r42,r43,r44,r45,r46,r47,r48,r49,r50,r51,r52}. 
In particular, the interplay of aperiodic order and non-Hermiticity has been investigated in several recent works \cite{r40,r41,r42,r43,r44,r45,r46,r47,r48,r49,r50,r51,r52}, revealing that the phase transition of eigenstates, from exponentially localized to extended (under periodic boundary conditions), can be often related to the change of topological (winding) numbers of the energy spectrum \cite{r31,r42,r50}.
  However, the dynamical behavior of the system near the phase transition, probed by the diffusion exponent or propagation speed of excitation, remains so far largely unexplored.\\
In this work we consider a non-Hermitian extension of the Aubry-Andr\'e-Harper model, where non-Hermiticity is introduced by asymmetric (non-recirpocal) hopping amplitudes  like in the Hatano-Nelson model \cite{r22,r23,r24,r31}. Special focus is devoted to the limit of unidirectional hopping, where rigorous analytical results can be obtained. The analysis unveils distinct features of dynamical phase transition when the hopping is asymmetric: while in the Hermitian case the the dynamical localization-delocalization transition is discontinuous in the diffusion exponent $\delta$ solely [Fig.1(a)], in the non-Hermitian case the phase transition is discontinuous both in the exponent $\delta$ {\em and} in the speed $v$ of ballistic transport [Fig.1(b)]. This means that, rather counter intuitively, ballistic transport with a finite speed is allowed in the lattice even very close to the critical point. Such a surprising result is related to the different spectral measure of the absolutely continuous spectrum near the phase transition point, which vanishes for symmetric hopping (Hermitian case) but not for asymmetric hopping (non-Hermitian case). Also, for a sufficiently large asymmetry in hopping amplitudes, we show that an increase of the potential $V$ results in an enhancement (rather than attenuation) of wave spreading in the lattice, thus providing an example of disorder-enhanced transport.\\
\begin{figure}[htbp]
  \includegraphics[width=86mm]{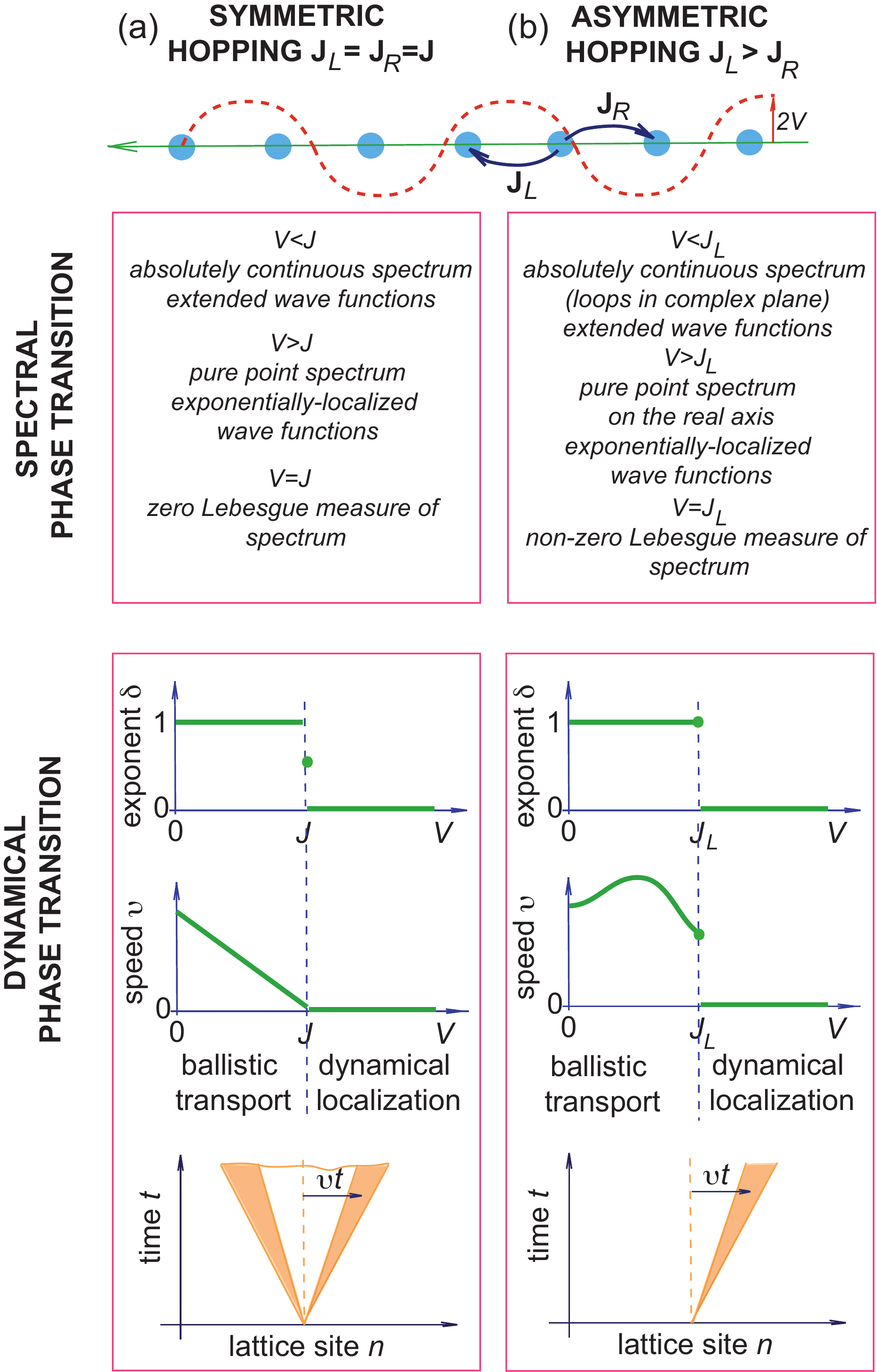}\\
   \caption{(color online) Schematic illustration of the main features associated to the spectral and dynamical phase transitions in the Aubry-Andr\'e-Harper model, as the potential amplitude $V$ is varied, for (a) symmetric (Hermitian), and (b) asymmetric (non-Hermitian) hopping. The bottom panels show a schematic of the wave spreading along the lattice, in the delocalized phase, on the space-time plane. Propagation is bidirectional for symmetric hopping, resulting in a full light cone pattern, while it is unidirectional for asymmetric hopping, resulting in a half light cone pattern.}
\end{figure}
\section{Dynamical phase transition in the Aubry-Andr\'e-Harper model: general results}
We consider the non-Hermitian extension of the Aubry-Andr\'e-Harper (AAH) model defined by the Hamiltonian
\begin{equation}
i \frac{d \psi_n}{dt}= J_R \psi_{n+1}+J_L \psi_{n-1} + 2 V \cos (2 \pi \alpha n) \psi_n \equiv H \psi_n
\end{equation} 
where $J_{L}$, $J_R$ are the left and right hopping amplitudes on the tight-binding lattice, $V $ is the amplitude of the on-site incommensurate potential, and $\alpha$ is irrational.
Without loss of generality, we assume $0 \leq J_R \leq J_L$ and $V \geq 0$. \stef{Owing to the non-Hermitian nature of the Hamiltonian, the dynamics described by Eq.(1) does not preserve the norm. This typically occurs when dealing with the dynamics of classical systems, such as photonic, acoustic, mechanical, or electrical  systems, where the non-unitary dynamics simply indicates that the system exchanges energy with the surrounding environment. On the other hand,  in an open quantum system we require, after any infinitesimal time step $dt$, to normalize the wave function, i.e. the state vector $ | \psi(t) \rangle$ of the system evolves according to the two-step process 
$ | \psi(t+dt) \rangle= \exp(-i H dt) | \psi (t) \rangle$ and $| \psi(t+dt \rangle)= | \psi(t+dt \rangle) / \|  \psi(t+dt \rangle )\|$. This two-time step procedure physically corresponds to the dynamics of postselected quantum trajectories in open quantum systems under continuous measurements, where quantum jumps are neglected \cite{r31,r35,figa1,figa2}.}\\
In this section we provide numerical results and some qualitative physical insights that highlight the different behavior of dynamical phase transitions 
in the Hemitian (symmetric hopping $J_L=J_R$) versus non Hermitian (asymmetric hopping $J_L>J_R$) models. The special case $J_R=0$, corresponding to unidirectional hopping on the lattice, is treated separately in Sec.III. Figure 1 illustrates the main features of spectral and dynamical phase transitions in Hermitian (symmetric hopping) and non-Hermitian (asymmetric hopping) AAH models.
\subsection{Symmetric hopping (Hermitian lattice)}
The Hermitian limit $J_R=J_L=J$ corresponds to the usual AAH model, and the Hamiltonian $H$ is referred to as the almost Mathieu operator in the mathematical literature. In this case the spectral and dynamical properties of $H$ are well understood \cite{R4,R6,r13,r20,r21a,r21b}. For irrational $\alpha$, the energy spectrum is a Cantor set and, for almost every $\alpha$ (e.g. with Diophantine properties), one has purely absolutely continuous
spectrum with extended states for $V<J$, purely singular continuous spectrum with critical wave functions for $V=J$, and pure
point spectrum with exponentially decaying wave functions for $V>J$. In the localized phase, all wave functions have the same localization length $\xi =1/L$, where $L$ is the energy-independent Lypaunov exponent given by
\begin{equation}
L= \log \frac{V}{J}.
\end{equation}
 The dynamical behavior of the system is captured by the asymptotic behavior at long times $t$ of the second-order moment of position operator  $\sigma^2(t)$  describing wave spreading, 
\begin{equation}
\sigma^2(t)= \frac{\sum_nn^2 |\psi_n(t)|^2}{\sum_n |\psi_n(t)|^2}
\end{equation} 
 with typical initial state  localized at the site $n=0$ in the lattice, i.e. $\psi_n(0)= \delta_{n.0}$. The asymptotic spreading of $\sigma^2(t)$ is described by a power law, i.e. $\sigma^2(t) \sim t^{2 \delta}$ where $\delta$ is dubbed the diffusion exponent.
 In the delocalized phase $V<J$, transport in the lattice is ballistic with the exponent $\delta(V)=1$. The excitation propagates bidirectionally along the lattice with  the non-vanishing speed 
\begin{equation}
v(V) \sim \frac{{\sigma(t)}}{t}
\end{equation}
which determines a full light cone pattern, as schematically shown in the bottom panel of Fig.1(a).
\begin{figure*}[htbp]
  \includegraphics[width=170mm]{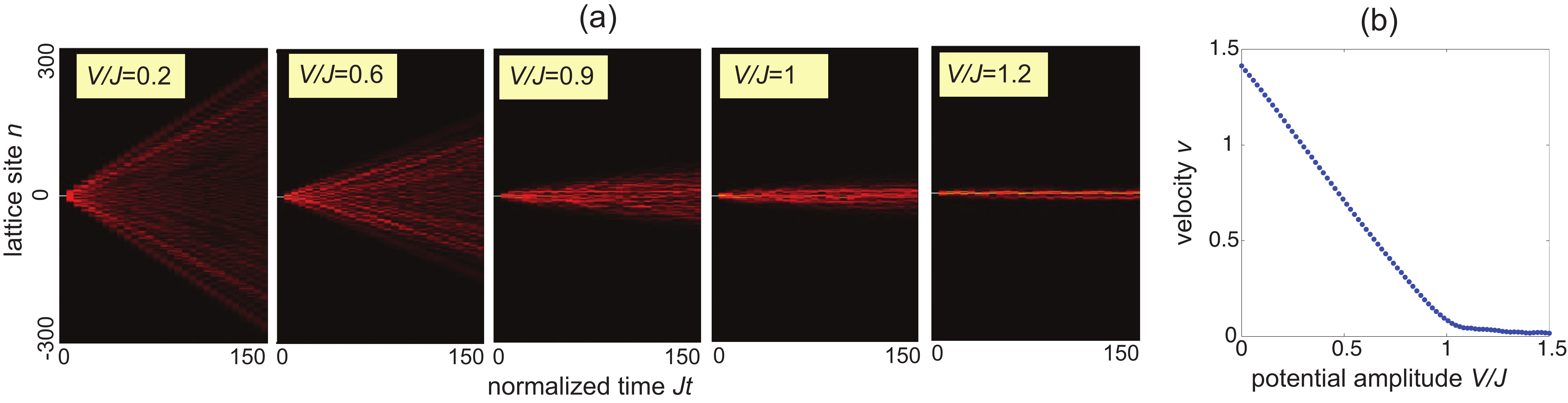}\\
   \caption{(color online) (a) Numerically-computed evolution of occupation amplitudes $|\psi_n(t)|$, for single-site input excitation $\psi_n(0)=\delta_{n,0}$, in the symmetric hopping case $J_L=J_R=J$ for a few increasing values of potential amplitude $V$ and for $\alpha=(\sqrt{5}-1)/2$. (b) Behavior of the spreading velocity $v(t)= \sigma(t)/t$ versus potential amplitude $V$, computed for the largest propagation time $t=150/J$. The spreading velocity $v$ is expressed in units of $J$.}
\end{figure*}
\begin{figure*}[htbp]
  \includegraphics[width=170mm]{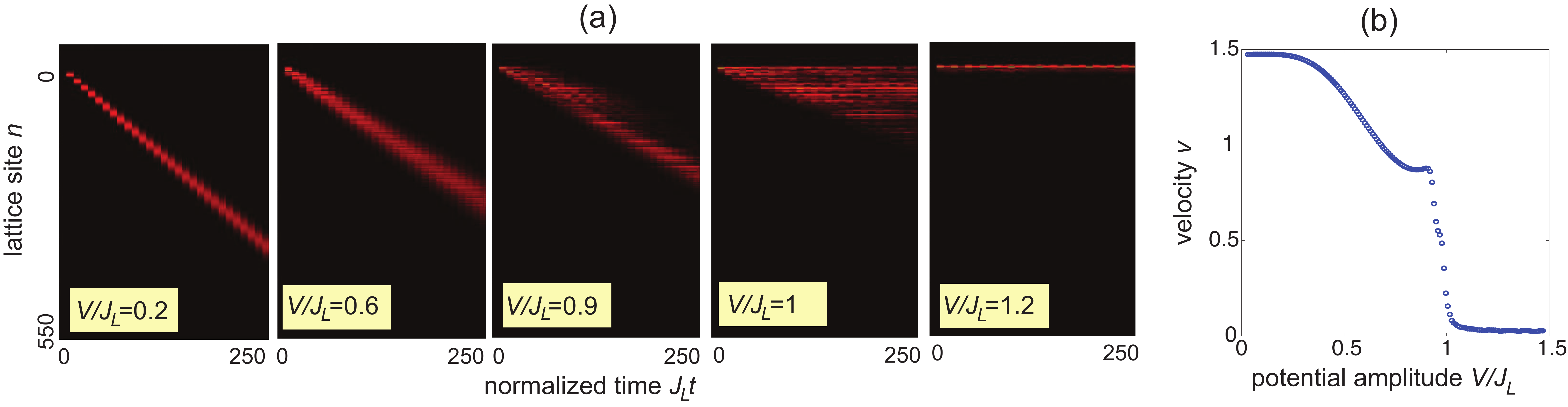}\\
   \caption{(color online) Same as Fig.2, but for asymmetric hopping with $J_R/J_L=0.5$. In the spreading dynamics, the amplitudes $\psi_n(t)$ have been normalized at each time step to $\sqrt{\sum_n | \psi_n(t)|^2}$. The spreading velocity $v$ is expressed in units of $J_L$.}
\end{figure*}
\begin{figure*}[htbp]
  \includegraphics[width=170mm]{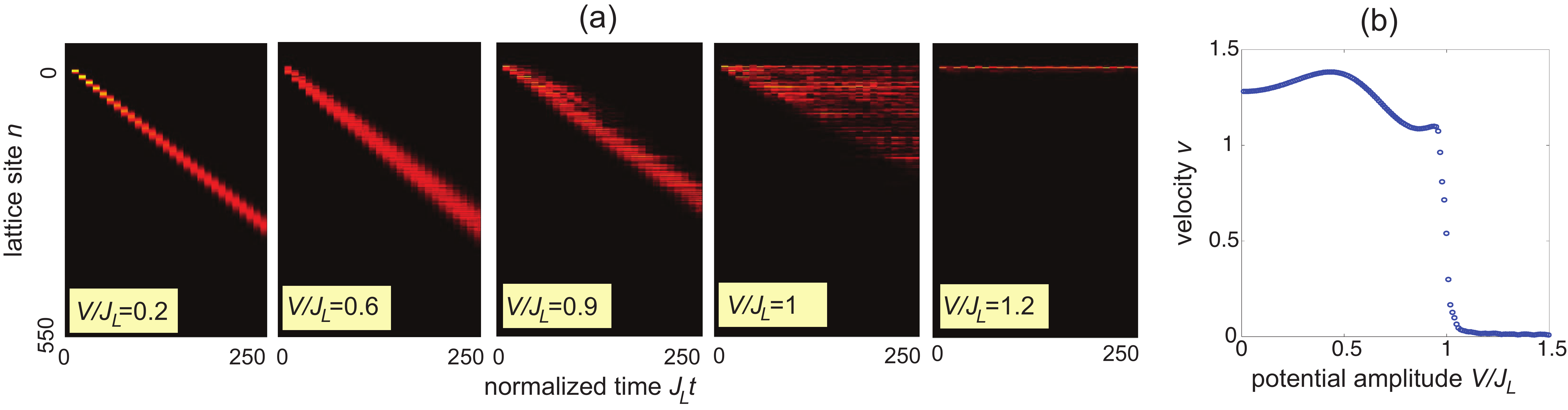}\\
   \caption{(color online) Same as Fig.3, but for $J_R/J_L=0.3$.}
\end{figure*}
On the other hand, in the localized phase $V>J$ transport is prevented (dynamical localization), so that $\delta(V)=v(V)=0$. At the phase transition point $V=V_c=J$, transport is intermediate between ballistic and localized; previous works have shown that transport in the lattice is nearly diffusive with an exponent $\delta(V_c) \simeq 0.5$ \cite{r13,r16,r53}. This means that the phase transition is characterized by a discontinuous behavior of diffusion exponent $\delta(V)$ near the critical point $V=V_c$ [Fig.1(a)]. \stef{ However, when one considers the speed $v(V)$ as an order parameter, the transition from ballistic to diffusive transport and localization is a smooth (continuous) process: $v(V)$ decreases almost linearly with $V$ and vanishes for $V \geq V_c$, with a discontinuity of the first derivative $(dv/dV)$ at $V=V_c$ This behavior is characteristic for a second-order phase transition }[Fig.1(a)]. Examples of numerically-computed wave spreading dynamics in the lattice for increasing values of $V/J$, and corresponding behavior of velocity $v(V)$, are shown in Fig.2, clearly suggesting that the phase transition is first-order in $v$. \stef{The numerical results are obtained by solving the time-domain Schr\"odinger equation (1) using a variable-step fourth-order Runge-Kutta method assuming single-site excitation at initial time; a sufficiently large number of lattice sites (typically 600 sites) has been considered to avoid edge effects up to the largest integration time.}\\
To physically understand why the ballistic speed $v$ diminishes and vanishes as the critical point is approached, let us consider a rational approximation $\alpha \sim q/p$ of the irrational $\alpha$, with $p,q$ prime integers, so that the actual incommensurate potential is approximated by a superlattice with period $p$ \cite{R6}. For example, for the inverse of the golden ratio $\alpha=( \sqrt{5}-1)/2=0.618033...$, the sequence $\alpha_n= q_n /p_n$ converges to $\alpha$ in the $n \rightarrow \infty$ limit, where $p_n=0,1,1,2,3,5,8,13,21,34,55, 89, 144, ...$ are the Fibonacci numbers and $q_n=p_{n-1}$. 
The Bloch wave functions of the superlattice satisfy the periodicity condition $\psi_{n+p}=\psi_n \exp(ikp)$, where $-\pi/p \leq k < \pi/p$  is the Bloch wave number, and the energy spectrum of $H$ is thus approximated by a set of $p$ energy bands 
with dispersion curves $E_l(k)$ ($l=1,2,3,...,p$), separated by $(p-1)$ energy gaps, obtained as solutions of a determinantal equation (see Appendix A). It is well known that, for large $p$, the sum $\Delta W$ of the widths $\Delta W_{l}$ of the allowed energy bands, i.e. the Lebesgue measure $\Delta W=\sum_l \Delta W_l$ of energy spectrum, is given by $\Delta W=4|J-V|$, vanishing as the critical point is attained \cite{R6}.  The corresponding Bloch eigenstates turn out to be delocalized over the entire period $p$ of the superlattice for $V<J$, whereas they tend to be tightly localized inside the superlattice period for $V>J$, with a localization length $ \xi=1/L= 1/ \log (V/J)$. To estimate the spreading behavior of the wave packet in the delocalized phase, we expand the initial state at time $t=0$ as a superposition of Bloch eigenstates of various bands, and make the (rather crude) approximation that each band is equally excited. The various wave packets in different bands propagate independently each other, and in the long time limit the wave packet displaying the largest spread is the one belonging to the band $l=l_0$ with the largest bandwidth $\Delta W_{l_0}$, which propagates at the largest group velocity $v_g$ given by
\stef{
 \[ 
 v_g \sim \frac{p \Delta W_{l_0}}{ 2}. \]
  
 Such a relation for the group velocity is justified as follows. The dispersion curve  $E_{l}(k)$ of a tiny superlattice band, with $k$ varying in the range $(-\pi /p, \pi / p$), can be approximated by the tight-binding curve $E_{l}(k)= (\Delta W_{l} /2 ) \cos (kp)$, where $\Delta W_{l}$ is the  full width of the band. The group velocity $v_g$ can be then estimated from the standard relation $v_g = |(d E_{l} / dk)_{\pi/2p}|$,  relating the excitation speed and band dispersion curve, i.e. $v_g \simeq p \Delta W_l / 2$.}
 Hence, at large times one expects the wave packet to spread far away from the initial site $n=0$ by the distance $\sigma(t) \sim (1/p)v_g t$, where the factor $(1/p)$ accounts for the excitation fraction of the $l_0$-th band. The wave packet thus undergoes ballistic spreading with a velocity $v(V) \sim (v_g/p) \sim \Delta W_{l_0} /2$. Taking into account that $\Delta W_{l_0} \leq \Delta W=4|J-V|$, one thus has $v(V) < \sim 2 |J-V|$, indicating that the ballistic speed $v(V)$ should vanish as $V$ approaches the critical value $V_c=J$. Note that, if all bands in the superlattice had the same bandwidth and were equally excited, they all would contribute to the asymptotic wave spreading and the above reasoning would give $v(V)  \sim \Delta W /2=2|J-V|$, corresponding to a linear decrease of $v$ with potential amplitude $V$ till to vanish at $V=J$, as observed in numerical simulations [see Fig.2(b)].
\subsection{Asymmetric hopping (non-Hermitian lattice)}
Let us now assume $J_R<J_L$, with a non-vanishing hopping $J_R>0$. The special case of unidirectional hopping, i.e. $J_R=0$, will be considered in more details in Sec.III. 
The spectral properties of the non-Hermitian AAH model with asymmetric hopping amplitudes have been studied in some recent works \cite{
r42,r50}. A localization-delocalization transition is found at the critical value $V=V_c$ of the on-site potential given by
\begin{equation}
V_c=J_L
\end{equation}
with all eigenstates extended and complex energy spectrum under periodic boundary conditions for $V<V_c$ (delocalized phase), and all eigenstates exponentially localized with real and pure-point spectrum for $V>V_c$ (localized phase). Interestingly, in the localized phase the energy spectrum of $H$ is the same as the one of the associated Hermitian AAH Hamiltonian $H_1$
\begin{equation}
H_1 \phi_n= J(\phi_{n+1}+\phi_{n-1})+ 2 V \cos (2 \pi \alpha n) \phi_n 
\end{equation}
with symmetric hopping amplitude $J$ given by
\begin{equation}
J= \sqrt{J_R J_L},
\end{equation}
while the eigenfunctions $\psi_n$ of $H$ are obtained from those $\phi_n$ of $H_1$ after multiplication by the term $\sim \exp(hn)$, i.e. $\psi_n= \exp(nh) \phi_n$, where we have set
\begin{equation}
h= \frac{1}{2} \log \left( \frac{J_L}{J_R}  \right).
\end{equation}
Hence, in the localized phase the localization lengths of the wave functions $\psi_n$ are asymmetric for left and right sides, which is reminiscent of the non-Hermitian skin effect \cite{skin1,skin2,skin3,skin4} in lattices with asymmetric hopping under open boundary conditions. 
A distinctive feature of the spectral phase transition between Hermitian and non-Hermitian models is that in the latter case the Lebesgue measure $\Delta W$ of energy spectrum  at the critical point $V=V_c=J_L$ does not vanish and reads
\begin{equation}
\Delta W_c=4 |J_L- \sqrt{J_R J_L}|.
\end{equation}
As we are going to discuss below, a non vanishing spectral measure at the phase transition point enables ballistic transport in the lattice in the delocalized phase with a velocity $v(V)$ which does not vanish as $V \rightarrow V_c^-$. 
\begin{figure}[htbp]
  \includegraphics[width=86mm]{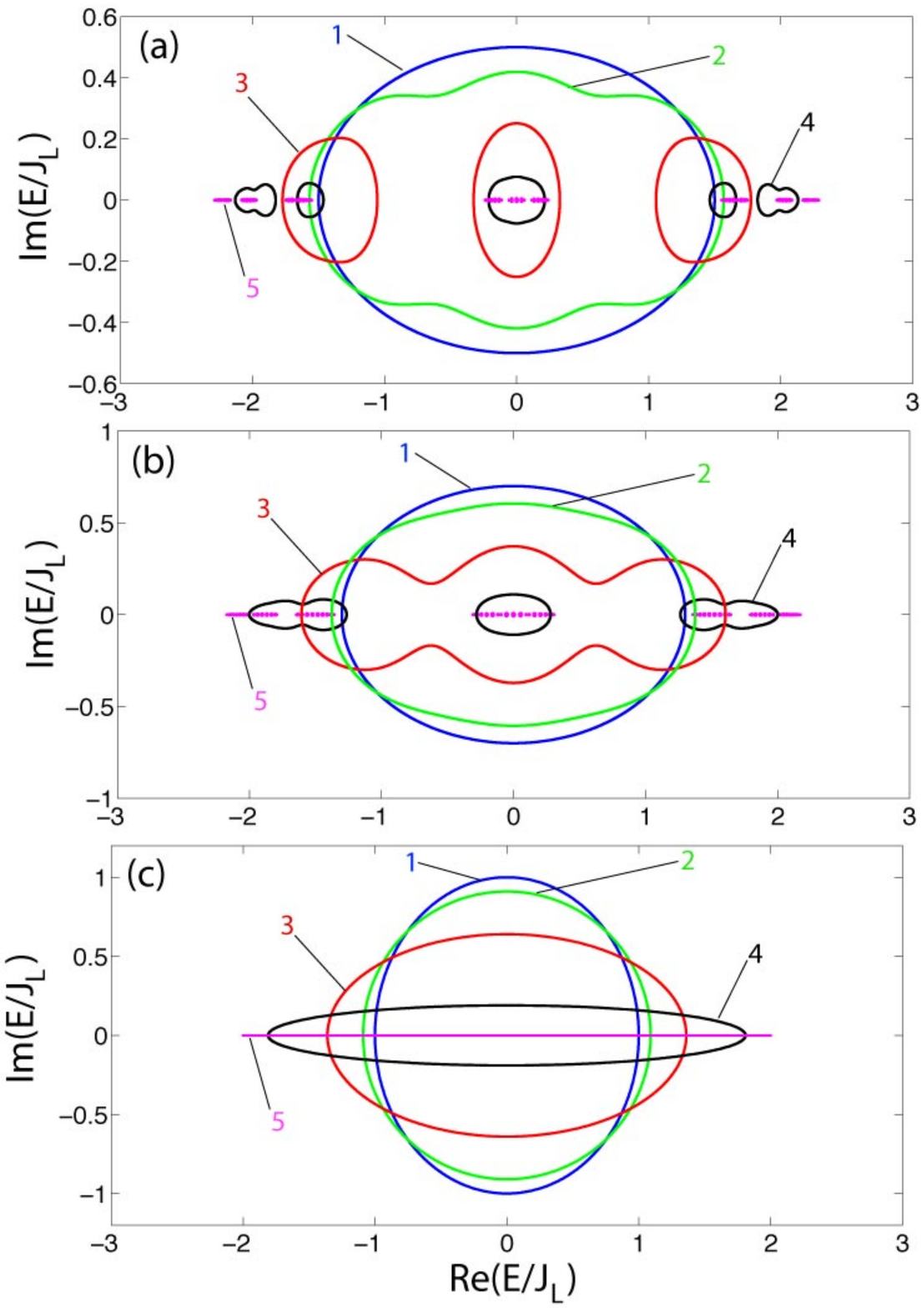}\\
   \caption{(color online) Energy spectrum in complex energy plane for the AAH Hamiltonian with asymmetric hopping and for a few increasing values of the on-site amplitude $V$. Curve 1: $V/J_L=0$; curve 2: $V/J_L=0.3$; curve 3: $V/J_L=0.6$; curve 4: $V/J_L=0.9$; curve 5: $V/J_L=1$ (phase transition point). In (a) $J_R/J_L=0.5$, in (b) $J_R/J_L=0.3$, whereas $J_R=0$  in (c) (unidirectional hopping(. The energy spectra in (a) and (b) have been numerically computed as the eigenvalues of the matrix $\mathcal{M}$ defined by Eq.(A5) in Appendix A for the rational approximation $q/p=89/144$ of the inverse of the golden ratio.}
\end{figure}
Figures 3 and 4 show a few numerical results of wave packet spreading in a non-Hermitian lattice and the numerically-computed behavior of the ballistic velocity $v(V)$ for two values of the ratio $J_R/J_L$. 
Note that, as the hopping is asymmetric, transport in the lattice is unidirectional \cite{Sci,Song}, so that in the delocalized phase excitation spreading is described by a half light cone.
The behavior of $v(V)$ versus the amplitude $V$ of on-site incommensurate potential unveils two major distinctive and somehow unexpected features:\\ 
(i) For $J_R/J_L$ smaller than $ \sim 0.5$, as $V$ is increased from zero the ballistic speed $v$ first increases (rather than decreases), i.e. wave spreading in the lattice becomes {\em faster} at larger disorder [see Fig.4(b)]. \stef{This result is rather counterintuitive and provides a noteworthy example of disorder-enhanced propagation \cite{cogl1,cogl2,cogl3,cogl4,Segev,cogl5,cogl6,cogl7,cogl8,cogl9} in the non-Hermitian realm}.\\
(ii) The ballistic speed  $v$ is discontinuous at the phase transition point $V=V_c$, i.e. excitation can propagate along the lattice at a non-vanishing velocity even arbitrarily close to the critical point.\\
To explain such a behavior, as in previous subsection let us consider the rational approximation $ \alpha \sim q/p$ of $\alpha$, with $p$, $q$ irreducible integer numbers, in the large $p$ limit. Let us indicate by $E_l(k)$  the dispersion curve of the $l$-th band of the superlattice described by the Hermitian Hamiltonian $H_1$, defined by Eq.(6). It can be readily shown that the $p$ energy bands of $H$, under periodic boundary conditions, are given by $E_l(k+ih)$, i.e they are obtained from the dispersion curves of the associated Hermitian lattice [Eq.(6)] after the replacement $k \rightarrow k+ih$, i.e. after complexification of the Bloch wave number, with $h=(1/2) \log (J_L/J_R)$ (technical details are given in Appendix A). \stef{It should be noted that in the current literature on the non-Hermitian skin effect  complexification of the Bloch wave number $k$, known as the generalized Brillouin zone, is relevant to determine the energy spectrum of a given Hamiltonian $H$ under the open boundary conditions in the thermodynamic limit \cite{skin1,skin3}: the energy spectrum of $H$ under open boundary conditions is obtained from the Hamiltonian $H(k)$ in Bloch space after complexification of $k$, so as $\beta \equiv \exp(ik)$ does not describe anymore a unit circle in complex plane. However, in our case complexification of $k$ is not related to the open boundary condition case: we consider {\em two} different superlattice models with Hamiltonians $H$ and $H_1$ in physical space, the former non-Hermitian and the latter Hermitian, and show that the energy spectra (Bloch mini bands) under periodic boundary conditions  of $H$  are obtained from those of $H_1$, under the same periodic boundary conditions, after complexification of $k$.} 
  For a vanishing potential amplitude $V=0$, we can take $p=1$, so that the Hermitian lattice of Eq.(6) displays the single tight-binding band $E_1(k)=2J \cos k$; correspondingly, the energy spectrum of the non-Hermitian lattice with asymmetric hopping reads 
\begin{equation}
E_1(k)=2 \sqrt{J_R J_L} \cos (k+ih)= J_R \exp(ik)+J_L \exp(-ik)
\end{equation}
describing an ellipse in the complex energy plane [see curve 1 in Figs.5(a) and (b)]. The largest velocity at which an excitation propagates along the lattice is given by the group velocity $v_g={\rm Re} \{ (dE/dk)_{k_m} \}$ at the Bloch wave number $k_m$ where the imaginary part of $E(k)$ takes its largest value \cite{Sci,probing}, i.e. $k_m=-\pi/2$ and $v_g=J_R+J_L$ for $V=0$. 
As the potential amplitude $V$ is increased, the energy spectrum in complex plane undergoes a smooth deformation from an ellipse, as shown by curve 2 in Figs.5(a) and (b). The deformation of the ellipse changes the dispersion relation and, for sufficiently small $J_R/J_L$, can result in an increase of the group velocity $v_g$, above the value $(J_L+J_R)$ found at $V=0$. Numerical results indicate that this happens for $J_R/J_L < \sim 0.5$, i.e it requires strong enough asymmetry between left and right hopping amplitudes. As the potential amplitude $V$ is further increased, the closed energy loop in complex energy plane splits into multiple separated loops; see Figs.5(a) and (b). The number of splitted loops increases as the critical point is approached, and their radius shrinks so as the energy spectrum becomes entirely real, with a fractal structure, at the critical point $V_c=J_L$ [curve 5 in Figs.5(a) and (b)]. For $V \geq V_c$, i.e. in the localized phase, the energy spectra of non-Hermitian $H$ and associated Hermitian $H_1$ Hamiltonians do coincide.  The group velocity $v_g$ that describes propagation of an excitation along the lattice in the delocalized phase close to the critical point $V_c=J_L$ can be estimated as $v_g \simeq p \Delta W_{l_0}  /2$, where $\Delta W_{l_0}$ is the width (real part of the energy) of the band of the superlattice displaying the largest value of imaginary part (thus dominating the dynamics at long times). A numerical  inspection of the band structure near the critical point indicates that the band with the largest imaginary part of energy also corresponds to the wider band in the real part of energy. Clearly, since $\sum_l \Delta W_l  \simeq \Delta W_c>0$ near the critical point, one has $\Delta W_{l_0} \geq \Delta W_c/p $ and thus $v_g \geq \sim \Delta W_c/2$, indicating that the speed of ballistic motion remains finite as the critical point is approached.

\section{Dynamical phase transition in the non-Hermitian Aubry-Andr\'e-Harper model with unidirectional hopping}
In this section we consider the special case corresponding to the AAH model with unidirectional hopping, i.e. $J_L=J>0$ and $J_R=0$. The Hamiltonian reads
\begin{equation}
H \psi_n= J \psi_{n-1}+2 V \cos ( 2 \pi \alpha n) \psi_n.
\end{equation}
\subsection{Energy spectrum and localization-delocalization transition}
The energy spectrum of this model can be determined in an exact form, as shown in Appendix B (see also \cite{r43}). The main results can be summarized as follows:\\
(i) For $V<V_c \equiv J$, the energy spectrum  $E$ is absolutely continuous and describes an ellipse in complex energy plane defined by the dispersion relation
\begin{equation}
E( \omega)=\left( J+ \frac{V^2}{J} \right) \cos \omega -i \left( J- \frac{V^2}{J} \right) \sin \omega
\end{equation}
with $-\pi \leq \omega < \pi$. The corresponding wave functions $\psi_n(\omega)$ are extended (generalized eigenfunctions). Note that the ellipse shrinks into a segment on the real axis, from $E=-2J$ to $E=2J$, as $V$ approaches the critical value $J$ [Fig.5(c)].\\
(ii) For $V \geq J$, the energy spectrum  $E$ is real, pure point and dense in the interval $(-2V,2V)$, with one-sided wave functions exponentially-localized with energy-independent localization length given by $\xi= 1/ \log (V / J)$.\\
Note that, compared to the lattice model with asymmetric but bidirectional hopping discussed in Sec. II.B, as the on-site potential amplitude $V$ is increased to approach the critical value $V_c$, the energy spectrum remains an ellipse and does not split into a set of loops [compare Figs.5(a,b) and Fig.5(c)]. Also, in the localized phase the spectrum is pure point but does not show the typical Cantor set structure of infinitely many small bands separated by  small gaps (the energy spectrum is the entire $(-2V,2V)$ interval). The reason thereof is that, while in the asymmetric hopping case with $J_R>0$ the energy spectrum of $H$ in the localized phase is the same as the one of the associated Hermitian AAH Hamiltonian $H_1$, thus displaying a fractal structure, in the unidirectional hopping limit $J_R=0$ such a correspondence becomes invalid and, as shown in Appendix B, the pure point energy spectrum is equidistributed in the full range $(-2V,2V)$.
\begin{figure*}[htbp]
  \includegraphics[width=170mm]{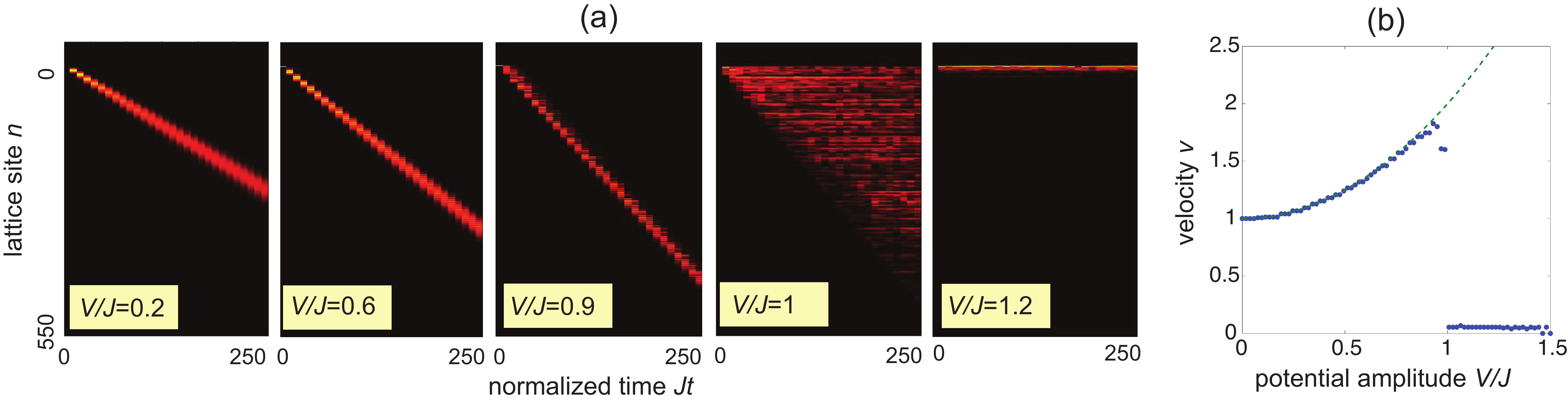}\\
   \caption{(color online) Same as Fig.3, but for $J_R=0$, $J_L=J$. The dashed line in (b) shows the behavior of velocity $v$ predicted by the saddle point method [Eq.(25)].}
\end{figure*}
\subsection{Dynamical behavior}
In this subsection we study the temporal behavior of a wave packet in the lattice with unidirectional hopping, and derive analytical expression of the largest velocity $v$ for propagation of excitation along the lattice that defines the half light cone aperture of Fig.1(b). Clearly, for $V>J$ the energy spectrum is real, all wave functions are exponentially localized with the same localization length, so that spectral localization implies also dynamical localization \cite{Simon}. Therefore, for $V>J$ one has $v(V)=0$ \stef{ and $\delta(V)=0$}. On the other hand, for $V<J$ (delocalized phase) the spectrum is absolutely continuous and transport in the lattice is expected to be ballistic ($\delta=1$) with largest propagation velocity $v=v(V)$, that we wish to calculate analytically. The most general wave packet at $t=0$ can be decomposed as a superposition of generalized eigenfunctions $\psi_n(\omega)$ of $H$ by suitable spectral amplitudes $F(\omega)$, which depend on the initial state. The solution to the Schr\"odinger equation $ i (d \psi_n /dt)=H \psi_n$ at times $t \geq 0$ reads
\begin{equation}
\psi_n(t)= \int_{-\pi}^{\pi} d \omega F( \omega) \psi_n( \omega) \exp[-i E(\omega) t]
\end{equation}
  where the dispersion relation $E=E(\omega)$ is given by Eq.(12) and where the form of generalized eigenfunctions $\psi_n(\omega)$ is given in Appendix B. To determine the largest propagation speed $v=v(V)$ of the wave packet along the lattice, we follow the method outlined in Ref.\cite{probing}, considering the asymptotic behavior at long times $t$ of the wave function along the space-time line $n=\nu t$, with some fixed velocity $\nu>0$, i.e. we consider the asymptotic behavior of
  \begin{equation}
  \psi(t) \equiv \psi_{n=v \nu }(t)=\int_{-\pi}^{\pi} d \omega F(\omega) \psi_{n= \nu t} (\omega) \exp[-i E(\omega) t]
  \end{equation}
  as $t \rightarrow \infty$. For $n>0$, the wave function $\psi_n(\omega)$ has the form [see Eq.(B2) in Appendix B]
  \begin{equation}
\psi_n (\omega)=  \prod_{l=1}^{n}  \frac{J}{E(\omega)-2 V \cos (2 \pi \alpha l)} 
\end{equation}
where we assumed, without loss of generality,  $ \psi_0(\omega)=1$. We can formally write
\begin{equation}
\psi_{\nu t}(\omega)= \exp[i \nu t S(\omega)]
\end{equation}
where we have set
\begin{equation}
S(\omega)= \frac{i}{ \nu t} \sum_{l=1}^{\nu t} \log \left(  \frac{E(\omega)-2 V \cos (2 \pi \alpha l)}{J} \right).
\end{equation}
In the large $\nu t$ limit, the sum on the right hand side of Eq.(17) can be approximated by an integral owing to the Weyl$^{\prime}$s equidistribution theorem and the properties of irrational rotations, namely one has
\begin{equation}
S(\omega)=\frac{i}{2 \pi} \int_{-\pi}^{\pi} dk \log \left(  \frac{E(\omega)-2 V \cos k}{J} \right) = {i} \gamma (\omega)
\end{equation}
where $\gamma(\omega)= -i \omega$ is the right Lyapunov exponent associated to the wave function $\psi_n(\omega)$ (see Appendix B for technical details). Hence, in the large $t$ limit we may assume $\psi_{\nu t}(\omega) \simeq \exp (i \nu t \omega)$ and thus
\begin{equation}
\psi(t) \sim \int_{-\pi}^{\pi} d \omega F(\omega) \exp [i \nu \omega t -i E(\omega)t].
\end{equation}
The growth rate $\lambda(\nu)$ of the wave packet along the space-time line $n= \nu t$, given by 
\begin{equation}
\lambda(\nu)= \lim_{t \rightarrow \infty} \frac{\log | \psi(t)|}{t} ,
\end{equation}
can be finally calculated by the saddle-point method \cite{probing}. This yields 
\begin{equation}
\lambda(v)=- \nu {\rm Im}( \omega_s)+ {\rm Im}(E(\omega_s))
\end{equation}
where the saddle point $\omega=\omega_s$ is the root of the equation
\begin{equation}
\left( \frac{dE}{d \omega} \right)_{\omega_s}=\nu
\end{equation}
in complex plane.
In our model, since $E(\omega)=2V \cos (\omega+i \rho)$, with $\rho=\log(J/V)$, it readily follows that $\omega_s=-i \rho- {\rm arcsin}( \nu /2V)$ and
\begin{equation}
\lambda( \nu)=\left\{
\begin{array}{ll}
\rho \nu  &  \nu <2V \\
\rho \nu- \nu {\rm arccosh} \left( \frac{\nu}{2V}\right)+ \sqrt{\nu^2-4V^2} & \nu>2V.
\end{array}
\right.
\end{equation}
The largest value $\lambda_m$ of the growth rate $\lambda(\nu)$ is given by 
\begin{equation}
\lambda_m=2 V \sinh \rho=J-\frac{V^2}{J}
\end{equation}
and it is attained at the speed $\nu=v$, given by
\begin{equation}
v=2 V \cosh \rho= J+\frac{V^2}{J}
\end{equation}
with the saddle point $\omega_s= -\pi/2$ on the real axis. The speed $v$ associated to the largest growth rate provides the aperture of the half light cone of Fig.1(b).
Equation (25) clearly shows that such a velocity is an increasing function of potential amplitude $V$, varying form $v=J$ at $V=0$ to $v=2J$ as the critical point $V=J$ is approached from below. Clearly, for $V>J$ one has $v=0$ (dynamical localization), thus proving that the behavior of $v(V)$ is discontinuous at the phase transition point $V=J$ \stef{(first-order phase transition)}.  Our analytical results have been confirmed by numerical simulations of wave packet spreading in the lattice, which are illustrated in Fig.6. In particular, the numerically-computed behavior of the velocity $v$, computed by the relation $v={\sigma(t)} /t$, turns out to be in very good agreement with Eq.(25) predicted by the saddle-point method.

\section{Conclusion}
The Aubry-Andr\'e-Harper model is the simplest and most studied one-dimensional model of aperiodic order displaying a delocalization-localization transition as the potential amplitude is increased above a finite threshold value. The abrupt transition in the energy spectrum (spectral phase transition), from absolutely continuous spectrum with extended wave functions in the delocalized phase to pure point spectrum with exponentially-localized wave functions in the localized phase, is associated to a distinct dynamical behavior of wave spreading in the lattice (dynamical phase transition), with ballistic transport in the delocalized phase and dynamical localization (suppression of wave spreading) in the localized phase. In terms of the velocity $v$ of wave spreading assumed as an order parameter, the dynamical phase transition \stef{ is of second order, i.e. $v=v(V)$ is a continuous function of the potential amplitude $V$, vanishes in the localized phase and its first derivative is discontinuous at the critical point $V=V_c$}. Similar spectral phase transitions have been recently found in certain non-Hermitian extensions of the Aubry-Andr\'e-Harper model \cite{r42,r43,r47,r48,r50}, however  the features of associated dynamical phase transition have been overlooked. In this work we unveiled distinct physical behavior in wave spreading and dynamical phase transitions in a non-Hermitian Aubry-Andr\'e-Harper model with asymmetric hopping amplitudes, as compared to the Hermitian limit of symmetric hopping. Remarkably, we found that for sufficiently strong asymmetry in the hopping  amplitudes the propagation of an excitation along the lattice is enhanced (rather than inhibited) by disorder.  Also, the dynamical phase transition \stef{is of first-order} in the velocity $v$, since $v(V)$ turns out to be discontinuous at the critical point. Such results provide important advances to understand the nontrivial interplay between disorder and non-Hermiticity, which is currently a hot area of research \cite{r29,r30,r31,r32,r33,r34,r35,r36,r38,r39,r42,r45,r47,r49,r51,r52}. \stef{The kind of non-Hermitian Hamiltonian with asymmetric hopping amplitudes, considered in this work, could be realized in synthetic matter using, for example, photonic systems \cite{r29,r42,Referee0,Referee1}, topoelectrical circuits \cite{Referee2}, mechanical metamaterials \cite{Referee3} or  in continuously-measured ultracold atom systems with reservoir engineering  \cite{r31,Referee4}.}

\acknowledgments
The author acknowledges the Spanish
State Research Agency through the Severo Ochoa and María
de Maeztu Program for Centers and Units of Excellence in
R\&D (Grant No. MDM-2017-0711).

\appendix
\section{Energy spectrum for a commensurate potential}
For rational $\alpha=q/p$, i.e. for a commensurate potential $V_n= 2 V \cos (2 \pi \alpha n)$, the Hamiltonian $H$ of the lattice,
\begin{equation}
 H \psi_n= J_R \psi_{n+1}+J_L \psi_{n-1}+V_n \psi_n,
\end{equation}
describes a superlattice of period $p$. Under periodic boundary conditions, its energy spectrum is thus absolutely continuous and is formed by a set of $p$ energy bands. According to the Bloch theorem, the eigenfunctions $\psi_n$ of $H$
satisfy the  Bloch condition
\begin{equation}
\psi_{n+p}=\psi_n \exp(ikp),
\end{equation}
where the Bloch wave number $k$ varies in the range $(-\pi/p, \pi/p)$, and can be assumed as a continuous variable for an infinitely extended lattice.
After setting $A_1=\psi_1$, $A_2=\psi_2$, ..., $A_p= \psi_p$, it then readily follows that the eigenvalue equation
\begin{equation}
 J_R \psi_{n+1}+J_L \psi_{n-1}+V_n \psi_n=E \psi_n
\end{equation}
is satisfied provided that
\begin{equation}
E \mathbf{A}= \mathcal{M} \mathbf{A}
\end{equation}
where we have set $\mathbf{A}=(A_1,A_2,...,A_p)^T$ and $\mathcal{M}$ is the $p \times p$ matrix given by
\begin{equation}
\mathcal{M}= 
\left(
\begin{array}{ccccccc}
V_1 & J_R & 0 & ... & 0 & 0 & J_L \exp(-ikp) \\
J_L & V_2 & J_R & ... & 0 & 0 & 0 \\
... & ... & ... & ... & ... & ... & ... \\
0 & 0 & 0 & ... & J_L & V_{p-1} & J_R \\
J_R \exp(ikp) & 0 & 0 & ... & 0 & J_L & V_p
\end{array}
\right).
\end{equation}
The energy bands $E=E_l(k)$ ($l=1,2,3,...,p$) are thus the $p$ eigenvalues of the matrix $\mathcal{M}$, which depend continuously on the Bloch wave number $k$.
Clearly, for symmetric hopping $J_R=J_L=J$ the Hamiltonian $H$ is Hermitian and the energies $E_l(k)$ of various bands are real. In the large $p$ limit, a large set of narrow bands separated by small gaps is found, which gives a Cantor set in the $p \rightarrow \infty$ limit.
The Lebesgue measure $\Delta W$ of energy spectrum is defined as the sum of the widths $\Delta W_l$ of various bands, i.e. $\Delta W= \sum_l \Delta W_l$, which converges toward $4|J-V|$ as $p \rightarrow \infty$ \cite{R6}.\\ 
For asymmetric hopping, the dispersion curves of energy bands take 
values in complex plane and describe rather generally one or more closed loops in complex energy plane (see Fig.5).
Interestingly, the dispersion curves $E_l(k)$ of the non-Hermitian lattice can be formally obtained from the one of an associated Hermitian lattice after complexification of the Bloch wave number $k$. In fact, let us consider the $p \times p$ diagonal matrix
\begin{equation}
\Lambda=
\left(
\begin{array}{ccccc}
1 & 0 & 0 & ... &0 \\
0 & \exp(-h) & 0 & ... & 0 \\
... & ... & ... & ... & ... \\
0 & 0 & 0 & ... & \exp(-ph+h)
\end{array}
\right)
\end{equation}
and the matrix $\mathcal{M}_1$, obtained from $\mathcal{M}$ by the similarity transformation
\begin{equation}
\mathcal{M}_1=\Lambda \mathcal{M} \Lambda^{-1}. 
\end{equation}
Clearly, for any Bloch wave number $k$ the matrices $\mathcal{M}$ and $\mathcal{M}_1$ have the same eigenvalues. If we assume $h= (1/2) \log (J_L / J_R)$, from Eqs.(A5), (A6) and (A7) it readily follows that
\begin{equation}
\mathcal{M}_1= 
\left(
\begin{array}{ccccccc}
V_1 & J & 0 & ... & 0 & 0 & J \exp(-i \kappa p) \\
J & V_2 & J & ... & 0 & 0 & 0 \\
... & ... & ... & ... & ... & ... & ... \\
0 & 0 & 0 & ... & J & V_{p-1} & J \\
J \exp(i \kappa p) & 0 & 0 & ... & 0 & J & V_p
\end{array}
\right)
\end{equation}
where we have set $J \equiv \sqrt{J_L J_R}$ and 
\begin{equation}
\kappa \equiv k+ih.
\end{equation}
 Equation (A8) clearly indicates that the energies $E_l$ of $H$ with asymmetric hopping $J_L> J_R$, i.e. the eigenvalues of $\mathcal{M}_1$,  are the same  
than the energies of an Hermitian Hamiltonian $H_1$ with symmetric hopping $J= \sqrt{J_L J_R}$ {\em but} with the Bloch wave number complexified according to Eq.(A9).

\section{Energy spectrum of the unidirectional Aubry-Andr\'e-Harper model}
The energy spectrum of the unidirectional Aubry-Andr\'e-Harper model is the set of complex numbers $E$ such that the solutions to the recurrence equation
\begin{equation}
E\psi_n= J \psi_{n-1}+V_n \psi_n ,
\end{equation}
with $V_n= 2 V \cos (2 \pi \alpha n)$, is not unbounded as $n \rightarrow \pm \infty$. To determine whether a given value $E$ belongs to the spectrum, we have to study the asymptomatic behavior of $\psi_n$ in the large $|n|$ limit. To this aim, let us distinguish two cases.\\ 
{\it First case.} Let us first consider a complex value of $E$ with $ E \notin (-2V,2V)$. Since $E \neq V_n$ for any $n$,
for a given value $\psi_0$ of the wave function amplitude at site $n=0$ from Eq.(B1) one has
\begin{equation}
\psi_n = \left( \prod_{l=1}^{n}  \frac{J}{E-V_l} \right)  \psi_0
\end{equation}
for $n>0$, and
\begin{equation}
\psi_{n} = \left( \prod_{l=-1}^{n}  \frac{E-V_{l+1}}{J} \right)  \psi_0
\end{equation}
for $n<0$. The right and left Lyapunov exponents $\gamma_{\pm}$, that determine the asymptotic behavior of $\psi_n$ as $n \rightarrow \pm \infty$, are given by
\begin{eqnarray}
\gamma_+ & = & -\lim_{n \rightarrow \infty} \frac{1}{n} \log \left( \frac{\psi_n}{\psi_0} \right) \\
\gamma_- & = & \lim_{n \rightarrow -\infty} \frac{1}{n} \log \left( \frac{\psi_n}{\psi_0} \right).
\end{eqnarray}
Using Eqs.(B2) and (B3), it readily follows that $\gamma_+=-\gamma_- \equiv \gamma$, with 
\begin{equation}
\gamma=\lim_{n \rightarrow \infty} \frac{1}{n} \sum_{l=1}^{n} \log \left(  \frac{E-2 V \cos (2 \pi \alpha l)}{J} \right).
\end{equation}
Note that, since $\gamma_-=-\gamma_+$, if the wave function were exponentially localized at $n \rightarrow \infty$, i.e. ${\rm Re}(\gamma_+)>0$, then it would be exponentially delocalized as $n \rightarrow -\infty$ since ${\rm Re}(\gamma_-)<0$, and viveversa. Hence the wave function cannot be exponentially localized for any value of $E$ outside the range ($-2V,2V)$, i.e. $E$ does not belong to the point spectrum of the Hamiltonain. However, provided that the real part of $\gamma$ vanishes the wave function is extended but does not secularly grow with $|n|$, i.e. $E$ belongs to the continuous spectrum of the Hamiltonian whenever ${\rm Re}(\gamma)=0$. 
For irrational $\alpha$, the limit on the right hand side of Eq.(B6) can be calculated using the Weyl$^{\prime}$s equidistribution theorem \cite{r43}, yielding 
\begin{equation}
\gamma=\frac{1}{2 \pi} \int_{- \pi}^{\pi} dk \log \left(  \frac{E-2 V \cos k}{J} \right).
\end{equation}
Since $E$ is not inside the interval $(-2V,2V)$, we may set $E= 2 V \cos \theta$, with $\theta$ a complex angle and ${\rm Im}(\theta)<0$. The integral on the right hand side of Eq.(B7) can be calculated in an exact form \cite{r43}, yielding the following final form for the Lyapunov exponent
\begin{equation}
\gamma= \log \frac{V}{J}+i \theta.
\end{equation}
Note that since ${\rm Im}(\theta)<0$, one has ${\rm Re}( \gamma)> \log (V/J)$. Therefore, for $V> J$ one has ${\rm Re }(\gamma)>0$, regardless of the value of energy $E$, so that $E$ does not belong to the continuous spectrum. On the other hand, for  $V \leq J$ the continuous spectrum is not empty and is composed by the set of complex energies $E$ such that  ${\rm Re}(\gamma)=0$, i.e. $\gamma=-i \omega$ with $\omega$ arbitrary real parameter. Using Eq.(B8), the condition $\gamma=-i \omega$ and Ansatz $E= 2V \cos \theta$, yields
\begin{eqnarray}
E & = & 2 V \cos \left(i \log \frac{V}{J} - \omega \right) \nonumber \\
& = &  J \exp(-i \omega)+ \frac{V^2}{J}  \exp(i \omega).
\end{eqnarray}
Note that, as $\omega$ varies in the range $(-\pi,\pi)$, the continuous spectrum $E$ describes an ellipse in complex energy plane, which shrinks toward the segment $(-2V,2V)$ on the real energy axis as $V$ approaches from below the critical value $V_c=J$ [Fig.5(c)].\\
{\it Second case.} Let us now assume that $E$ is real and inside the range $(-2V,2V)$. Precisely, let us assume that $E=2 V \cos( 2 \pi \alpha n_0)$ for some integer $n_0$. Note that, since $\alpha$ is irrational, the set of energies $E$ obtained when $n_0$ varies from $-\infty$ to $\infty$ is dense and equidistributed in the range $(-2V,2V)$. In this case the solution to Eq.(B1) reads explicitly
\begin{equation}
\psi_n=
\left\{
\begin{array}{ll}
0 & n < n_0 \\
1 & n=n_0 \\
\prod_{l=n_0+1}^{n} \frac{J}{E-V_l} & n>n_0
\end{array}
\right.
\end{equation}
The right Lyapunov exponent reads
\begin{equation}
\gamma_+=- \lim_{n \rightarrow \infty} \frac{1}{n} \log \psi_n= \frac{1}{2 \pi} \int_0^{2 \pi} dk \log \left(   \frac{E-2 V \cos k}{J} \right).
\end{equation}
For $E \in (-2V,2V)$, the integral on the right hand side of Eq.(B11) turns out to be independent of $E$ and equals to $\log(V/J)$, i.e.
\begin{equation}
\gamma_+= \log \left( \frac{V}{J} \right).
\end{equation}
For $V<J$, $\gamma_+<0$ and thus $\psi_n$ exponentially grows as $n \rightarrow \infty$, i.e. $E$ does not belong to the point spectrum of the Hamiltonian. On the other hand, for $V>J$ one has $\gamma_+>0$ 
for any energy $E=2 V \cos (2 \pi \alpha n_0)$, dense in the range $(-2V,2V)$: hence $E$ belongs to the point spectrum of the Hamiltonian. Since the Lyapunov exponent $\gamma_+$ does not depend on energy $E$,  the wave function is one-sided exponentially localized with an energy-independent localization length $\xi$ given by
\begin{equation}
\xi= \frac{1}{\gamma_+}= \frac{1}{\log (V/J)}.
\end{equation}
Finally, for $V=J$ one has $\gamma_{+}=0$, indicating that at the critical point the wave function is one-sided extended.

\end{document}